\documentclass[pra,tightenlines,nofootinbib,superscriptaddress,10pt,twocolumn,showpacs]{revtex4-1} %reprint, 

\usepackage[colorlinks=true,urlcolor=blue,citecolor=blue,linkcolor=blue,bookmarks=false]{hyperref} %
\usepackage{amsmath,amssymb,mathrsfs,amsfonts,bm}
\usepackage{graphicx,color}
\usepackage{extarrows}
\usepackage{dcolumn}
\usepackage{siunitx}

\newcommand{\degree}{^\circ}
\newcommand{\transpose}{\mathsf{T}}
\newcommand{\angstrom}{\textup{\AA}}
\newcommand{\tabincell}[2]{\begin{tabular}{@{}#1@{}}#2\end{tabular}}

\begin{document}

%-----------------------------------------------------------------
% documentation    title,  authors,   abstract,   pacs
%-----------------------------------------------------------------

\title {Visualizing Pairing Symmetry in Nematic Superconductors using spin-polarized spectroscopy of magnetic impurities}

\author{Liang Chen}
%\email[Corresponding Email: ]{slchern@ncepu.edu.cn}
\affiliation{School of Mathematics and Physics, North China Electric Power University, Beijing, 102206, China}
%\affiliation{Beijing Computational Science Research Centre, Beijing, 100084, China}

\author{Ya-Li Zhang}
\affiliation{School of Mathematics and Physics, North China Electric Power University, Beijing, 102206, China}

\author{Rong-Sheng Han}
\email[Corresponding Email: ]{hrs@ncepu.edu.cn}
\affiliation{School of Mathematics and Physics, North China Electric Power University, Beijing, 102206, China}

\date{\today}

\begin{abstract}
  We study the spin-polarized spectral properties of Yu-Shiba-Rusinov resonance states induced by magnetic impurities in 2- and 3-dimensional nematic superconductors: few layer Bi$_2$Te$_3$ grown on FeTe$_{0.55}$Se$_{0.45}$ (2-dimensional) and Cu$_x$Bi$_2$Se$_3$ (3-dimensional). We focus on the relationship between pairing symmetry and the topograph of spin-polarized spectroscopy. We calculate the spin-polarized local density of states (SP LDOS) and the corresponding Fourier transformation using the $T$-matrix method for both the 2- and 3-dimensional materials. Various situations with different impurity orientations and different SP LDOSs have been investigated. We find that, like the quasiparticle interference spectrum, the SP LDOS can be applied to distinguish other pairings which preserve the threefold rotation symmetry of $D_3$ point group and nematic pairings in these materials.
\end{abstract}


\maketitle

%-----------------------------------------------------------------
% The body of the paper
%-----------------------------------------------------------------

\section{Introduction} \label{sec1} %-----------------------------

The search for topological superconductors (TSCs) is an important topic in recent years \cite{QiXL2011RMP,SatoM2017RPP}. Nontrivial topology of the superconducting gap in TSC provides a buildable platform for the study of Majorana bound states, exotic quasiparticles which are their own antiparticles; obey Non-Abelian statistics; and have been considered as fundamental building blocks for future topological quantum computation \cite{NayakC2008RMP}. Unconventional superconductor Sr$_2$RuO$_4$ has been investigated for many years and been believed to be an intrinsic TSC with spin-triplet pairing \cite{MackenzieAP2003RMP}. Recent study on iron-based superconductor shows that the surface states of FeTe$_{0.55}$Se$_{0.45}$ are intrinsic topological superconducting \cite{ZhangP2018Science}. In addition to these prominent candidates, another series of materials are investigated in recent years as potential TSCs, Cu$_x$Bi$_2$Se$_3$ and its variants, Sr$_x$Bi$_2$Se$_3$, Nb$_x$Bi$_2$Se$_3$, Tl$_x$Bi$_2$Se$_3$, etc.

The parent material of these candidates, Bi$_2$Se$_3$, is a strong topological insulator with bulk band gap about $0.3$ eV and gapless surface states protected by time-reversal symmetry \cite{ZhangH2009NatPhys,LiuCX2010PRB}. When the Fermi energy is tuned into the conduction or valence band by chemical doping, strong spin-orbit coupling may drive these materials into a topological nontrivial superconducting phase at low temperature. Experimental investigations show that \cite{HorYS2010PRL}, if the copper atoms are intercalated in between the quintuple layers of Bi$_2$Se$_3$, the Fermi energy will be tuned into the conduction band, and there will be a superconducting transition at maximum $T_C=3.8$ K for copper concentration $0.12<x<0.15$. Theoretical investigations of Fu and Berg \cite{FuL2010PRL} show that, the order parameter of this superconductor should be parity-odd; belong to the 2-dimensional $E_u$ representation of the $D_3$ point group; and support topological nontrivial Andreev bound states. This kind of pairing potential, which preserves the twofold rotation symmetry but spontaneously breaks the threefold rotation symmetry, is generally named nematic pairing potential \cite{FuL2014PRB}. Experimental investigations \cite{SasakiS2011PRL,KirzhnerT2012PRB} show that there is an in gap zero-bias conductance peak (ZBCP) in the point contact spectra measurement on the surface of Cu$_x$Bi$_2$Se$_3$. The existence of the ZBCP confirms that Cu$_x$Bi$_2$Se$_3$ should be a parity-odd topological superconductor. However, detailed measurements based on scanning tunneling microscope (STM) \cite{LevyN2013PRL,DuG2017NatComms} and Andreev reflection spectroscopy \cite{PengH2013PRB} give negative evidences. It is reported that the ZBCP disappears for clean materials. This pronounced peak is more likely induced by Andreev reflection between the point contact and the superconductor. It is further suggested that the pairing potential is $s$-wave symmetric \cite{LevyN2013PRL}. On the other hand, various experiments support that threefold rotation symmetry is broken and that these materials are nematic pairing \cite{Fu2016NatPhys}, i.e., nuclear magnetic resonance \cite{MatanoK2016NatPhys,NagaiY2016PRB}, thermodynamics \cite{YonezawaS2016NatPhys}, upper critical field \cite{YonezawaS2016NatPhys,VenderbosJ2016PRB,PanY2016SciRep,ShenJ2017npjQM}, penetration depth \cite{SmylieMP2016PRB,SmylieMP2017PRB}, torque magnetometry \cite{AsabaT2017PRX}, superconducting gap versus in-plane magnetic field measurement \cite{TauR2018PRX}, magnetoresistance \cite{KuntsevichAY2018NJP}, and transport experiment \cite{DuG2017SCPMA}, etc. In addition to these 3-dimensional materials, recently few layer Bi$_2$Te$_3$ grown on FeTe$_{0.55}$Se$_{0.45}$ substrate is demonstrated to be nematic pairing by using the quasiparticle interference (QPI) and magnetic flux vortex measurements \cite{ChenM2018SciAdv}, etc. Despite the experimental evidences of nematic pairing listed above, different pairing potentials are proposed from theoretical considerations, e.g., momentum-dependent pairing potentials proposed in early stage \cite{ChenL2013JPCM}, chiral pairing \cite{WuF2017PRB} belongs to the $E_u$ representation of $D_3$ point group, etc.

The identification of pairing symmetry in Cu$_x$Bi$_2$Se$_3$ and its variants plays an critical role to distinguish the topological nontrivial nematic pairing and the trivial one. Experimentally, the phase-sensitive tetracrystal measurements \cite{TsueiCC1994PRL,TsueiCC1997Nature}, the spacial distribution of Yu-Shiba-Rusinov resonance states induced by magnetic impurities \cite{Yu1965APS,Shiba1968CSS,Rusinov1969JETPL}, and QPI spectrum  \cite{Hoffman2002Science,WangQH2003PRB,Hanaguri2007NatPhys,Hanaguri2009Science} are typical methods applied to identify the pairing symmetry of superconductors. In this work, we propose another method, the spin-polarized spectroscopy (SPS), to discern the nematic pairing. SPS of magnetic impurities in superconductors has been studied in previous works \cite{KaladzhyanV2016PRBa,KaladzhyanV2016PRBb} to measure the spin-orbital couplings in superconductors. Here we systematically study the SPS of magnetic impurities located on the surface of Cu$_x$Bi$_2$Se$_3$ and Bi$_2$Se$_3/$FeTe$_{0.55}$Se$_{0.45}$ superconductors. We find that, for these materials with strong spin-orbit coupling, the SPS can give a clear manifestation of the resonance state and help us to distinguish the pairing potentials which belong to different representations of the $D_3$ point group.

The rest of this article is organized as follows: theoretical investigations are presented in Sec. \ref{sec2}. Model Hamiltonian, the derivation of spin-polarized (SP) LDOS and QPI, and symmetry analysis of $D_3$ point group are given in this section. In Sec. \ref{sec3}, we show the numerical results of SP LDOS for both the 2- and 3-dimensional systems. A conclusion is given in Sec. \ref{sec4}.

\section{Theoretical Investigation}\label{sec2} %-------------------------

\subsection{Model Hamiltonian}
We use the following two-orbital tight binding models \cite{ZhangH2009NatPhys,FuL2009PRL,LiuCX2010PRB,HaoL2014PRB,HaoL2017PRB} to describe the normal state of Cu$_x$Bi$_2$Se$_3$ ($h_{3d}$) and Bi$_2$Te$_3$ grown on FeTe$_{0.55}$Se$_{0.45}$ ($h_{2d}$),
\begin{gather}
  \hat{H}=\sum_{\bm{k}}\psi^{\dag}_{\bm{k}}h(\bm{k})\psi_{\bm{k}}, \label{eq1} \\
  h_{3d}(\bm{k})=\varepsilon(\bm{k})\Gamma_0+M(\bm{k})\Gamma_5+B_0c_z\Gamma_4+A_0\left[c_y(\bm{k})\Gamma_1\right.\notag\\
  \left.-c_x(\bm{k})\Gamma_2\right]+R_1d_1(\bm{k})\Gamma_3+R_2d_2(\bm{k})\Gamma_4. \label{eq2} \\
  h_{2d}(\bm{k})=\varepsilon(\bm{k})\Gamma_0+M(\bm{k})\Gamma_5+A_0[c_y(\bm{k})\Gamma_1-c_x(\bm{k})\Gamma_2] \notag \\
  +R_1d_1(\bm{k})\Gamma_3+R_2d_2(\bm{k})\Gamma_4, \label{eq3}
\end{gather}
$\psi^{\dag}_{\bm{k}}$ and $\psi_{\bm{k}}$ are the creation and annihilation operators of electron states with wave-vector $\bm{k}$. The explicit expression of $\psi_{\bm{k}}$ is $(a_{\bm{k},\uparrow},a_{\bm{k},\downarrow},b_{\bm{k},\uparrow}, b_{\bm{k},\downarrow})^{\transpose}$, here $a$ and $b$ refer to, respectively, the two low-energy $p_z$ orbitals of Se atoms on the top and bottom layers of each Bi$_2$Se$_3$ quintuple unit. $\Gamma_0$ to $\Gamma_5$ are $4\times4$ matrices acting in the orbital and spin space. $\Gamma_0=\sigma_0\otimes{s_0}$ is the identity matrix, $\Gamma_1=\sigma_z\otimes{s_x}$, $\Gamma_2=\sigma_z\otimes{s_y}$,  $\Gamma_3=\sigma_z\otimes{s_z}$,  $\Gamma_4=-\sigma_y\otimes{s_0}$,  and $\Gamma_5=\sigma_x\otimes{s_0}$. $\sigma_{\#}$ and $s_{\#}$ (${\#}=x,y,z$) are Pauli matrices in the orbital and spin spaces, respectively.
%$\varepsilon_0(\bm{k})+M(\bm{k})$ and $\varepsilon_0(\bm{k})-M(\bm{k})$ represent the conventional band structures for the two $p_z$ orbitals.
$\varepsilon(\bm{k})$ represents the normal dispersion, the off-diagonal terms proportional to $M(\bm{k})$ represent the band inversion of topological insulator, the terms with $A_0$ and $B_0$ contained represent the in-plane and out of-plane spin-orbit couplings, the last two terms in Eqs. (\ref{eq2}) and (\ref{eq3}) induce the hexagonal warping effect of the Fermi surface. For completeness, here we list the details of these terms represented in tight-binding model. $\varepsilon(\bm{k})=C_0+2C_1[1-\cos(\bm{k}\cdot\bm{\delta}_4)]+\frac{4}{3}C_2[3-\cos(\bm{k}\cdot\bm{\delta}_1)-\cos(\bm{k}\cdot\bm{\delta}_2)-\cos(\bm{k}\cdot\bm{\delta}_3)]$, $M(\bm{k})=M_0+2M_1[1-\cos(\bm{k}\cdot\bm{\delta}_4)]+\frac{4}{3}M_2[3-\cos(\bm{k}\cdot\bm{\delta}_1)-\cos(\bm{k}\cdot\bm{\delta}_2)-\cos(\bm{k}\cdot\bm{\delta}_3)]$, $c_x(\bm{k})=\frac{1}{\sqrt{3}}[\sin(\bm{k}\cdot\bm{\delta}_1)-\sin(\bm{k}\cdot\bm{\delta}_2)]$, $c_y(\bm{k})=\frac{1}{3}[\sin(\bm{k}\cdot\bm{\delta}_1)+\sin(\bm{k}\cdot\bm{\delta}_2)-2\sin(\bm{k}\cdot\bm{\delta}_3)]$, $c_z(\bm{k})=\sin(\bm{k}\cdot\bm{\delta}_4)$, $d_1(\bm{k})=-\frac{8}{3\sqrt{3}}[\sin(\bm{k}\cdot\bm{a}_1)+\sin(\bm{k}\cdot\bm{a}_2)+\sin(\bm{k}\cdot\bm{a}_3)]$,  $d_2(\bm{k})=-8[\sin(\bm{k}\cdot\bm{\delta}_1)+\sin(\bm{k}\cdot\bm{\delta}_2)+\sin(\bm{k}\cdot\bm{\delta}_3)]$. Here $\bm{\delta}_j$ and $\bm{a}_j$ represent the nearest-neighboring and in-plane next nearest-neighboring bond vectors, respectively. Their detailed expressions are, $\bm{\delta}_1=(\frac{\sqrt{3}}{2}a,\frac{1}{2}a,0)$, $\bm{\delta}_2=(-\frac{\sqrt{3}}{2}a,\frac{1}{2}a,0)$, $\bm{\delta}_3=(0,-a,0)$, $\bm{\delta}_4=(0,0,c)$, $\bm{a}_1=\bm{\delta}_1-\bm{\delta}_2$, $\bm{a}_2=\bm{\delta}_2-\bm{\delta}_3$, $\bm{a}_3=\bm{\delta}_3-\bm{\delta}_1$. $a=4.14\angstrom$ and $3c=28.64\angstrom$ are the in-plane and out-of-plane lattice spacings of Cu$_x$Bi$_2$Se$_3$; for Bi$_2$Te$_3$ grown on FeTe$_{0.55}$Se$_{0.45}$ substrate, $a=4.38\angstrom$. The parameters used for numerical calculation are listed in Table \ref{tab1}. These Hamiltonians, Eqs. (\ref{eq2}) and (\ref{eq3}), preserve the desired rotation symmetry of $D_{3d}$ point group. $D_{3d}$ point group has 12 elements which are generated by 3 elementary operations: threefold rotation along the $z$-axis, twofold rotation along the $y$-axis, and inversion. The matrix representations of these elementary operations are given in Table \ref{tab2} for both the normal state and superconducting cases.

\begin{table}[tb]
	\caption{Parameters used for numerical calculation of the tight-binding model \cite{HaoL2014PRB,HaoL2017PRB}, in units of electron volts (eV).} \label{tab1}
	\centering
	\begin{tabular}{p{1.5cm}|p{1.2cm}p{1.2cm}p{1.2cm}p{1.2cm}p{1.2cm}}
		\hline\hline
		& $C_0$ & $C_1$ & $C_2$ & $M_0$ & $M_1$ \\
		\hline
		Cu$_x$Bi$_2$Se$_3$ & $-$0.008 & 0.06 & 1.0 & $-$0.26 & 0.3 \\
		Bi$_2$Te$_3$ & $-$0.18 & 0.0634 & 2.59 & $-$0.3 & 0.102 \\
		\hline
	\end{tabular}
	\begin{tabular}{p{1.5cm}|p{1.2cm}p{1.2cm}p{1.2cm}p{1.2cm}p{1.2cm}}
		\hline
		& $M_2$ & $A_0$ & $B_0$ & $R_1$ & $R_2$ \\
		\hline
		Cu$_x$Bi$_2$Se$_3$ & 1.2 & 0.8 & 0.35 & 0.2 & $-$0.3 \\
		Bi$_2$Te$_3$ & 2.991 & 0.655 & -- & 0.536 & $-$1.064  \\
		\hline\hline
	\end{tabular}
\end{table}

\begin{table}[tb]
	\caption{Matrix representation of the elementary operations for $D_{3d}$ point group. $\mathscr{C}_z$ refers to the threefold rotation operation along the $z$-axis. $\mathscr{C}^{\prime}_{y}$ refers to the twofold rotation operation along the $y$-axis. $\mathcal{I}$ is the inversion operation. Here $\zeta_0$ and $\zeta_z$ are the $2\times2$ identity matrix and the third Pauli matrix in the Nambu spinor space.} \label{tab2}
	\centering
	\begin{tabular}{c|ccc}
		\hline\hline
		operations & $\mathscr{C}_z$ & $\mathscr{C}^{\prime}_{y}$ & $\mathscr{I}$ \\
		\hline
		\tabincell{l}{matrix rep. \\ (normal state)}    & $\sigma_0\otimes{e}^{2\pi{i}s_z/3}$ & $i\sigma_0\otimes{s_x}$ & $\sigma_x\otimes{s_0}$ \\
		\hline
		\tabincell{l}{matrix rep. \\ (BdG)}    & $\sigma_0\otimes{e}^{2\pi{i}s_z\otimes{\zeta_z}/3}$ & $i\sigma_0\otimes{s_x}\otimes\zeta_z$ & $\sigma_x\otimes{s_0}\otimes\zeta_0$ \\
		\hline\hline
	\end{tabular}
\end{table}

For both the 3- and 2-dimensional situations, the superconducting Hamiltonian can be written in the Bogliubov-de Gennes (BdG) formalism as follows by using the Nambu spinor basis  $\Psi^{\dag}_{\bm{k}}=(\psi^{\dag}_{\bm{k}}, \psi_{-\bm{k}}^{\transpose})$,
\begin{gather}
\hat{\mathcal{H}}_{\textrm{BdG}}=\frac{1}{2}\sum_{\bm{k}}\Psi_{\bm{k}}^{\dag}\left(\begin{matrix}	
h(\bm{k})&\Delta \\
\Delta^{\dag}&-h^{\transpose}(-\bm{k})
\end{matrix}\right)\Psi_{\bm{k}}, \label{eq4}
\end{gather}
where $\Delta$ is the superconducting pairing potential, $h(\bm{k})$ equals to $h_{3d}(\bm{k})$ and $h_{2d}(\bm{k})$ for the 3- and 2-dimensional systems respectively. In this work, we focus on the pairing potentials belong to the 2-dimensional $E_u$ representation of the $D_{3d}$ point group (the nematic pairing). As shown in previous studies \cite{FuL2010PRL,HaoL2014PRB,HaoL2017PRB}, these pairing potentials in the $E_u$ representation can be written as, $\Delta_{4a}=i\Delta_a\sigma_y\otimes{s_0}$ and $\Delta_{4b}=\Delta_b\sigma_y\otimes{s_z}$, with $\Delta_a$ and $\Delta_b$ referring to the strength of pairing potentials. Generally, these pairing potentials break the threefold rotation symmetries of $D_{3d}$ point group.

\subsection{SP LDOS and QPI}
The retarded Green's function of the BdG Hamiltonian can be evaluated using the standard method, which gives,
\begin{equation}
	\mathcal{G}_{0}(E,\bm{k})=\left[E+i0^{+}-\left(\begin{matrix}
	h(\bm{k}) & \Delta \\  \Delta^{\dag} & -h^{\transpose}(-\bm{k})
	\end{matrix}\right)\right]^{-1}. \label{eq5}
\end{equation}
We assume that there is an impurity located at the original point, i.e., $\hat{V}=\delta(\bm{R})\psi^{\dag}(\bm{R})\mathcal{V}\psi(\bm{R})$. For nonmagnetic and magnetic impurities, the impurity potential take the following forms,
\begin{eqnarray}
&\mathcal{V}^{\rm NM}&=J_0{\sigma_0}\otimes{s_0},  \label{eq6}\\
\textrm{and,~~~} &\mathcal{V}^{\rm M}&={\sigma_0}\otimes\bm{J}\cdot\bm{s}, \label{eq7}
\end{eqnarray}
the superscripts, NM and M, refer to the nonmagnetic and magnetic impurity potentials. $J_0$ and $\bm{J}=(J_x, J_y, J_z)$ represent the strength of scattering potential for nonmagnetic and magnetic impurities, respectively. Here, we have assumed that the impurity potential is short-ranged.

In the following studies, we consider two different configurations: (1) for 3-dimensional material Cu$_x$Bi$_2$Se$_3$, described by the Hamiltonian (\ref{eq2}), the impurity is located on the surface of the material, i.e., the $xy$-plane at $z=0$;
(2) for few layer Bi$_2$Te$_3$ grown on FeTe$_{0.55}$Se$_{0.45}$ substrate, described by the 2-dimensional Hamiltonian (\ref{eq3}), the impurity is located at the 2-dimensional plane. For both these two cases, $\bm{R}=(x,y)$.
% the impurity is located on the surface of 3-dimensional superconductor, in this case, the surface is set to be the $xy$-plane at $z=0$ and the bulk Hamiltonian for the normal state is given in Eq. (\ref{eq2});

When the Kondo temperature is much smaller than the superconducting gap, the magnetic impurity is not screened by the itinerant electrons, and it can be considered as classical impurity \cite{BalatskyAV2006RMP}. In this scenario, the full Green's function of the electrons affected by the impurity can be solved by using the standard $T$-matrix method. For the short-ranged impurity potential given in Eqs. (\ref{eq6}) and (\ref{eq7}), the $T$-matrix is expressed as,
\begin{equation}
\mathcal{T}(E)=[\mathcal{V}^{-1}-\mathcal{G}_0(E,\bm{0})]^{-1}, \label{eq8}
\end{equation}
where $\mathcal{G}_0(E,\bm{0})$ is the free Green's function at $\bm{R}=(0,0)$. For the 2-dimensional Hamiltonian, the non-perturbed retarded Green's function, $\mathcal{G}_0(E,\bm{R})$ is evaluated as the Fourier transformation of Eq. (\ref{eq5}),
\begin{gather}
	\mathcal{G}_0(E,\bm{R})=\frac{1}{N}\sum_{\bm{k}}e^{i\bm{k}\cdot\bm{R}}\mathcal{G}_0(E,\bm{k}), \label{eq9}
\end{gather}
where $N$ is the number of wave-vectors in summation.
For the 3-dimensional Hamiltonian (the impurity is located at the original point on the surface $z=0$), $\mathcal{G}_0(E,\bm{R})$ is the non-perturbed retarded Green's function of surface states, which can be derived numerically by using the fast iterative method \cite{SanchoMPL1984JPFMP}. The full Green's function in real space is given by,
\begin{gather}
	\mathcal{G}(E,\bm{R})=\mathcal{G}_0(E,\bm{R})+\Delta\mathcal{G}(E,\bm{R}), \label{eq10} \\
	\Delta\mathcal{G}(E,\bm{R})=\mathcal{G}_0(E,\bm{R})\mathcal{T}(E)\mathcal{G}_0(E,-\bm{R}). \label{eq11}
\end{gather}

The SP LDOS is defined as,
\begin{gather}
S_z(E,\bm{R})=-\frac{1}{\pi}\textrm{Im}[\Delta\mathcal{G}_{11}-\Delta\mathcal{G}_{22}+\Delta\mathcal{G}_{33}-\Delta\mathcal{G}_{44}], \label{eq12} \\
S_x(E,\bm{R})=-\frac{1}{\pi}\textrm{Im}[\Delta\mathcal{G}_{12}+\Delta\mathcal{G}_{21}+\Delta\mathcal{G}_{34}+\Delta\mathcal{G}_{43}], \label{eq13} \\
S_y(E,\bm{R})=-\frac{1}{\pi}\textrm{Re}[\Delta\mathcal{G}_{12}-\Delta\mathcal{G}_{21}+\Delta\mathcal{G}_{34}-\Delta\mathcal{G}_{43}], \label{eq14}
\end{gather}
where $\Delta\mathcal{G}_{ij}$ takes the $i$th row and $j$th column component of the matrix $\Delta\mathcal{G}(E,\bm{R})$. Generally, the orientation of magnetic impurity and SP LDOS can take various combinations, e.g., $\bm{J}=(J,0,0)$ (the magnetic impurity is oriented in the $x$-direction) and $S_z(E,\bm{R})$ (the SP LDOS of $z$-component is detected).

In addition to the SP LDOS, the QPI is another important method to identify the pairing symmetry of superconductors, which is defined as follows for nonmagnetic impurity,
\begin{gather}
	\delta\rho(E,\bm{q})=\sum_{\bm{R}}e^{i\bm{q}\cdot\bm{R}}\delta\rho(E,\bm{R}), \label{eq15} \\
	\delta\rho(E,\bm{R})=-\frac{1}{\pi}\textrm{Im}[\Delta\mathcal{G}_{11}+\Delta\mathcal{G}_{22}+\Delta\mathcal{G}_{33}+\Delta\mathcal{G}_{44}]. \label{eq16}
\end{gather}
Here we need to notify that, generally, $\delta\rho(E,\bm{q})$ and the Fourier transformation of SP LDOS, $S_{\#}(E,\bm{q})=\int\mathrm{d}\bm{R}e^{-i\bm{q}\cdot\bm{R}}S_{\#}(E,\bm{q})$ ($\#=x,y,z$) should take complex numbers, though $\delta\rho(E,\bm{R})$ and $S_{\#}(E,\bm{R})$ are real functions.

\subsection{Symmetry analysis}
In this subsection, we analyze the symmetries of different SP LDOS patterns if the pairing potential is invariant under some operations of the $D_3$ point group. Generally, the normal state Hamiltonian preserves all the symmetries listed in Table \ref{tab2}. Due to the strong spin-orbit coupling in these materials, if the pairing potential breaks some symmetries listed in Table \ref{tab2}, the SP LDOS patterns should reveals these symmetry-breakings.
% Here we consider only the 2-dimensional situation.
% the distribution of the SP LDOS can be applied to trace the symmetry of pairing potentials.

For the 3-dimensional case, the situation is a bit complicated. We consider only the STM measurable resonance states, i.e., the impurity should be located on the surface of Cu$_x$Bi$_2$Se$_3$. In this case, the inversion and twofold rotation symmetry related to $z\rightarrow-z$ is broken. Only the threefold rotation symmetry may be survival. In the following symmetry analysis, we focus on the 2-dimensional situation. For both 2- and 3-dimensional situations, numerical investigations and analysis of SP LDOS will be presented in the next section.

Firstly, we consider the twofold rotation operation along the $y$-axis, $\mathscr{C}_y^{\prime}$ given in Table \ref{tab2}. Suppose a pairing potential $\Delta$ preserves this symmetry, i.e.,
\begin{equation}
	\mathscr{C}^{\prime-1}_y\left(\begin{matrix}
	0 & \Delta \\ \Delta^{\dag} & 0
	\end{matrix}\right)\mathscr{C}^{\prime}_y=\left(\begin{matrix}
	0 & \Delta \\ \Delta^{\dag} & 0
	\end{matrix}\right). \label{eq17}
\end{equation}
In this case, the BdG Hamiltonian should be invariant under this operation,
\begin{equation}
	\mathscr{C}_{y}^{\prime-1}h_{\rm BdG}(k_x, k_y)\mathscr{C}_{y}^{\prime}=h_{\rm BdG}(-k_x, k_y), \label{eq18}
\end{equation}
here we have used the explicit expression of wave vector $\bm{k}=(k_x,k_y)$, and
\begin{equation}
	h_{\rm BdG}(k_x, k_y)=\left(\begin{matrix}
	h(k_x, k_y) & \Delta \\  \Delta^{\dag} & -h^{\transpose}(-k_x, -k_y)
	\end{matrix}\right). \label{eq19}
\end{equation}
Using the definition of unperturbed Green's function, Eq. (\ref{eq5}), we get,
\begin{equation}
	\mathscr{C}_{y}^{\prime-1}\mathcal{G}_{0}(E, k_x, k_y)\mathscr{C}_{y}^{\prime}=\mathcal{G}_{0}(E, -k_x, k_y). \label{eq20}
\end{equation}
If the orientation of magnetic impurity takes the form $\bm{J}=(J,0,0)$, the impurity potential $\mathcal{V}$ should be invariant under the $\mathscr{C}^{\prime}_y$ operation, so we get $\mathscr{C}_{y}^{\prime-1}\mathcal{V}\mathscr{C}_{y}^{\prime}=\mathcal{V}$, and,
\begin{equation}
	\mathscr{C}_{y}^{\prime-1}\mathcal{T}(E)\mathscr{C}_{y}^{\prime}=\mathcal{T}(E). \label{eq21}
\end{equation}
Substituting Eqs. (\ref{eq20}) and (\ref{eq21}) into (\ref{eq9}) and (\ref{eq11}), we find,
\begin{equation}
  \mathscr{C}_{y}^{\prime-1}\Delta\mathcal{G}(E, x, y)\mathscr{C}_{y}^{\prime}=\Delta\mathcal{G}(E, -x, y). \label{eq22}
\end{equation}
where $(x,y)$ is the explicit expression of location $\bm{R}$. Using the definition of SP LDOSs given in Eqs. (\ref{eq12})-(\ref{eq14}), we can find that,
\begin{gather}
	S_{z}(E,x,y)=-S_{z}(E,-x,y), \label{eq23} \\
	S_{x}(E,x,y)=+S_{x}(E,-x,y), \label{eq24} \\
	S_{y}(E,x,y)=-S_{y}(E,-x,y). \label{eq25}
\end{gather}

If the pairing potential is invariant under the inversion operation $\mathscr{I}$ given in Table \ref{tab2}, after a similar derivation, we find,
\begin{equation}
\mathscr{I}\Delta\mathcal{G}_0(E,x,y)\mathscr{I}=\Delta\mathcal{G}(E,-x,-y), \label{eq26}
\end{equation}
and
\begin{gather}
	S_{z}(E,x,y)=+S_{z}(E,-x,-y), \label{eq27} \\
	S_{x}(E,x,y)=+S_{x}(E,-x,-y), \label{eq28} \\
	S_{y}(E,x,y)=+S_{y}(E,-x,-y). \label{eq29}
\end{gather}
for arbitrary impurity spin orientation $\bm{J}=(J_x,J_y,J_z)$.

Finally, if the pairing potential is invariant under the threefold rotation operation $\mathscr{C}_z$ given in Table \ref{tab2} and the spin of magnetic impurity is oriented at $z$-direction, $\bm{J}=(0,0,J)$, straightforward calculation gives,
\begin{equation}
\mathscr{C}_{z}^{-1}\Delta\mathcal{G}(E,\bm{R})\mathscr{C}_{z}=\Delta\mathcal{G}(E,\tilde{\bm{R}}).  \label{eq30}
\end{equation}
where $\tilde{\bm{R}}=(-\frac{1}{2}x-\frac{\sqrt{3}}{2}y,\frac{\sqrt{3}}{2}x-\frac{1}{2}y)$ for $\bm{R}=(x,y)$. Using the expressions of SP LDOS, we get,
\begin{gather}
	S_{z}(E,\bm{R})=S_z(E,\tilde{\bm{R}}), \label{eq31} \\
	\left(-\frac{1}{2}S_x+\frac{\sqrt{3}}{2}S_{y}\right)(E,\bm{R})=S_x(E,\tilde{\bm{R}}), \label{eq32} \\
	\left(-\frac{\sqrt{3}}{2}S_x-\frac{1}{2}S_{y}\right)(E,\bm{R})=S_y(E,\tilde{\bm{R}}). \label{eq33}
\end{gather}

\begin{figure*}[tb]
	\centering
	\includegraphics[width=\textwidth]{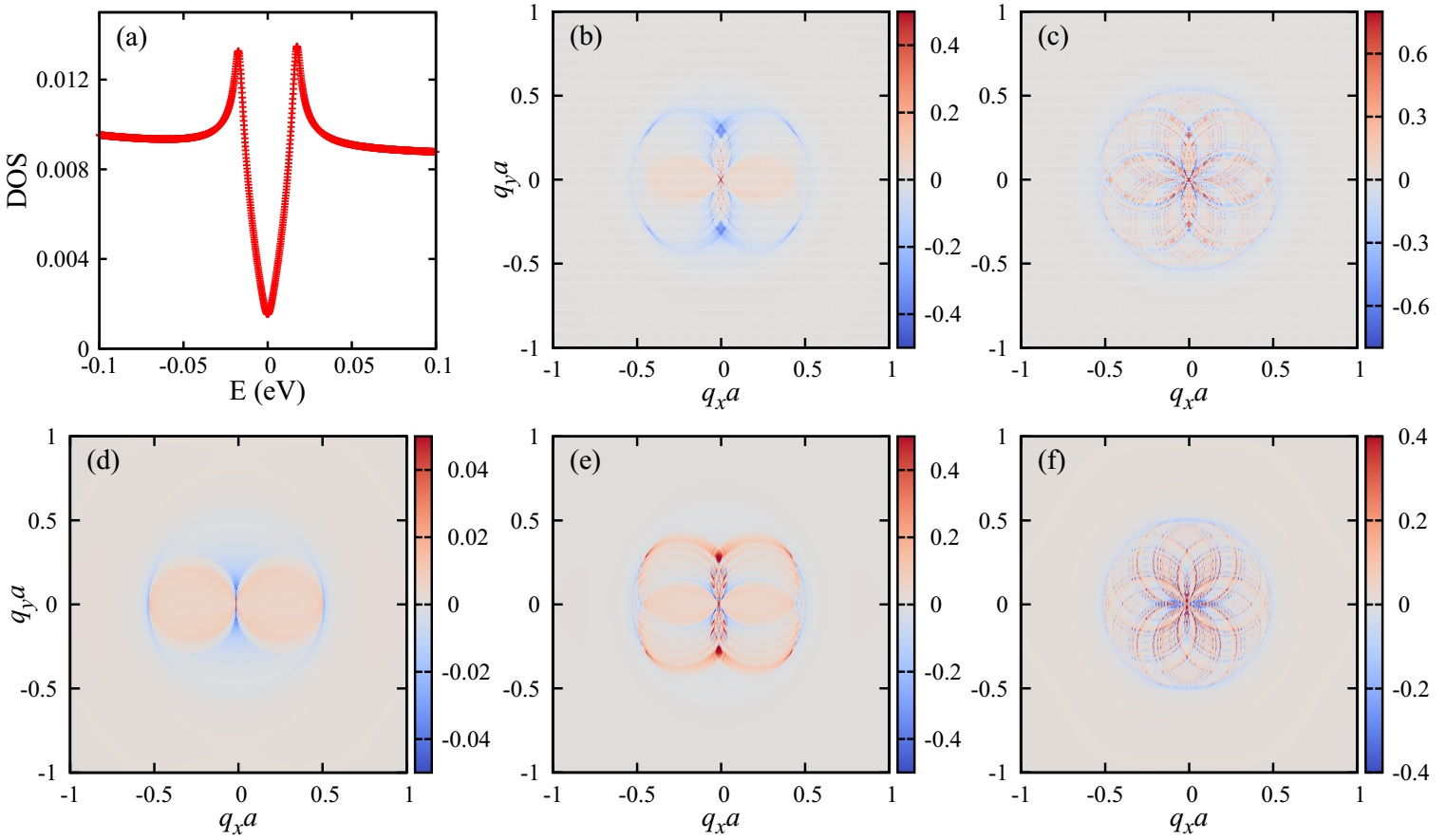}
	\caption{ QPIs of superconducting Bi$_2$Te$_3$ grown on FeTe$_{0.55}$Se$_{0.45}$ substrate with pairing potential $\Delta_{4a}$ and different energies. (a) DOS for the BdG Hamiltonian, Eq. (\ref{eq4}), with parameters given in Table \ref{tab1} (Bi$_2$Te$_3$). (b)-(f) Contour plots of QPI spectra $\delta\rho(E,\bm{q})$ defined in Eq. (\ref{eq15}) for different energies: $E=-0.01$eV, $-0.03$eV, $0$, $-0.01$eV, and $-0.03$eV, respectively. The Fermi energy $E_F=0.32$eV and pairing potential $\Delta_a=0.02$eV is chosen for numerical calculation. 
	}
	\label{fig1}
\end{figure*}

\begin{figure*}[tb]
	\centering
	\includegraphics[width=\textwidth]{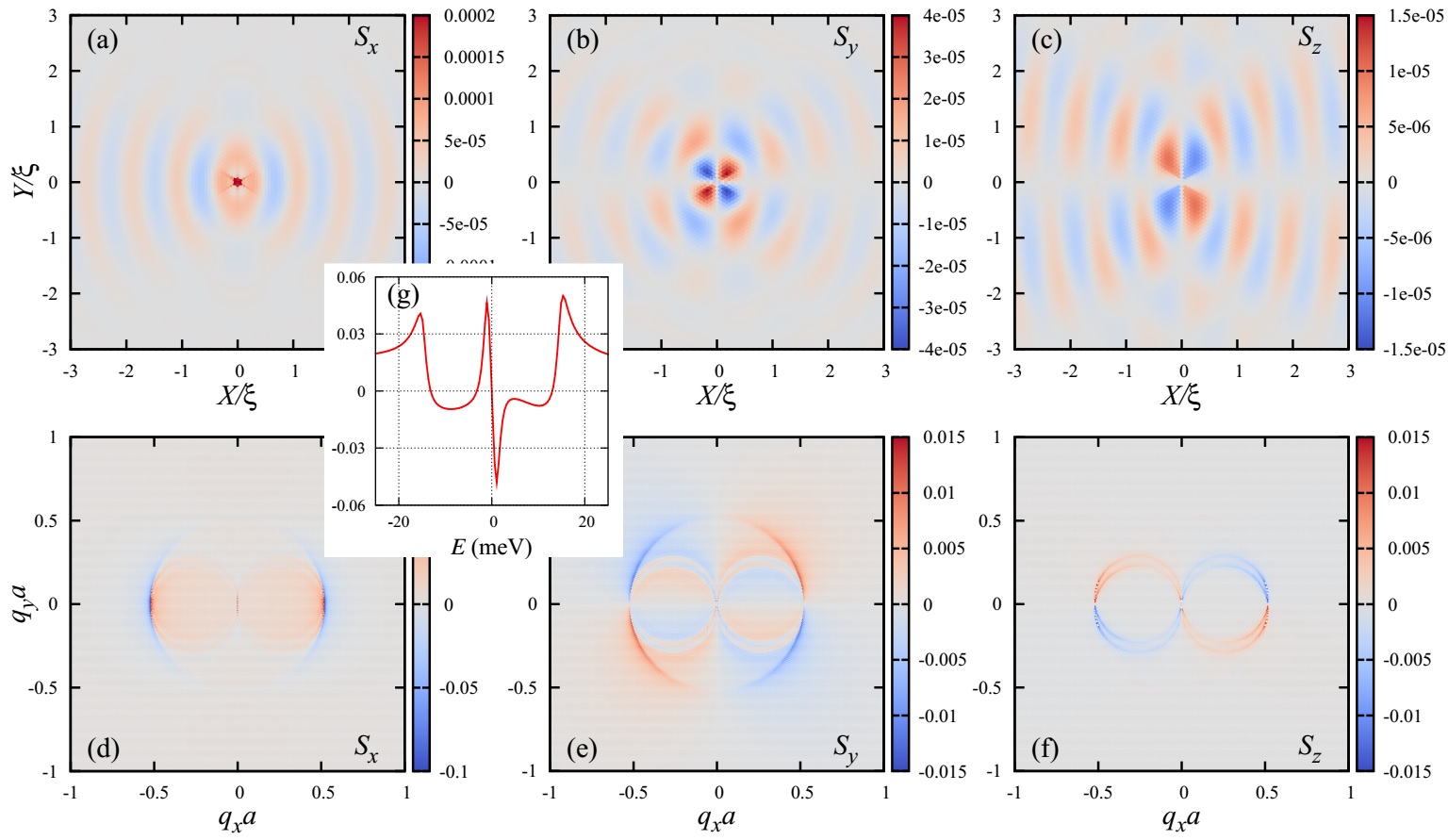}
	\caption{SP LDOSs of superconducting Bi$_2$Te$_3$ grown on FeTe$_{0.55}$Se$_{0.45}$ substrate with pairing potential $\Delta_{4a}$ and $x$-impurity, $\bm{J}=(J,0,0)$. Upper panel: SP LDOSs in real space, $S_{\#}(E,\bm{R})$ with $\#=x,y,z$. $\xi=\hbar{v_F}/\Delta$ is the coherence length, which is estimated to be about $15a$ ($a$ is the lattice spacing) for the given parameters shown in Table \ref{tab1} and superconducting gap $\Delta_{a}=0.02$eV. Lower panel: SP LDOS in wave-vector space, $S_{\#}(E,\bm{q})$ with $\#=x,y,z$. The insert show the integration of $S_{x}(E,\bm{R})$ over $\bm{R}$, $S_{x}(E,\bm{q}=0)$, versus energy $E$. In-gap resonance peaks are located at $E_{R}=\pm1.05$ meV. Impurity energy $J=1$eV is chosen for numerical calculation, other parameters are the same as shown in Fig. \ref{fig1}.
	}
	\label{fig2}
\end{figure*}

\begin{figure*}[tb]
	\centering
	\includegraphics[width=\textwidth]{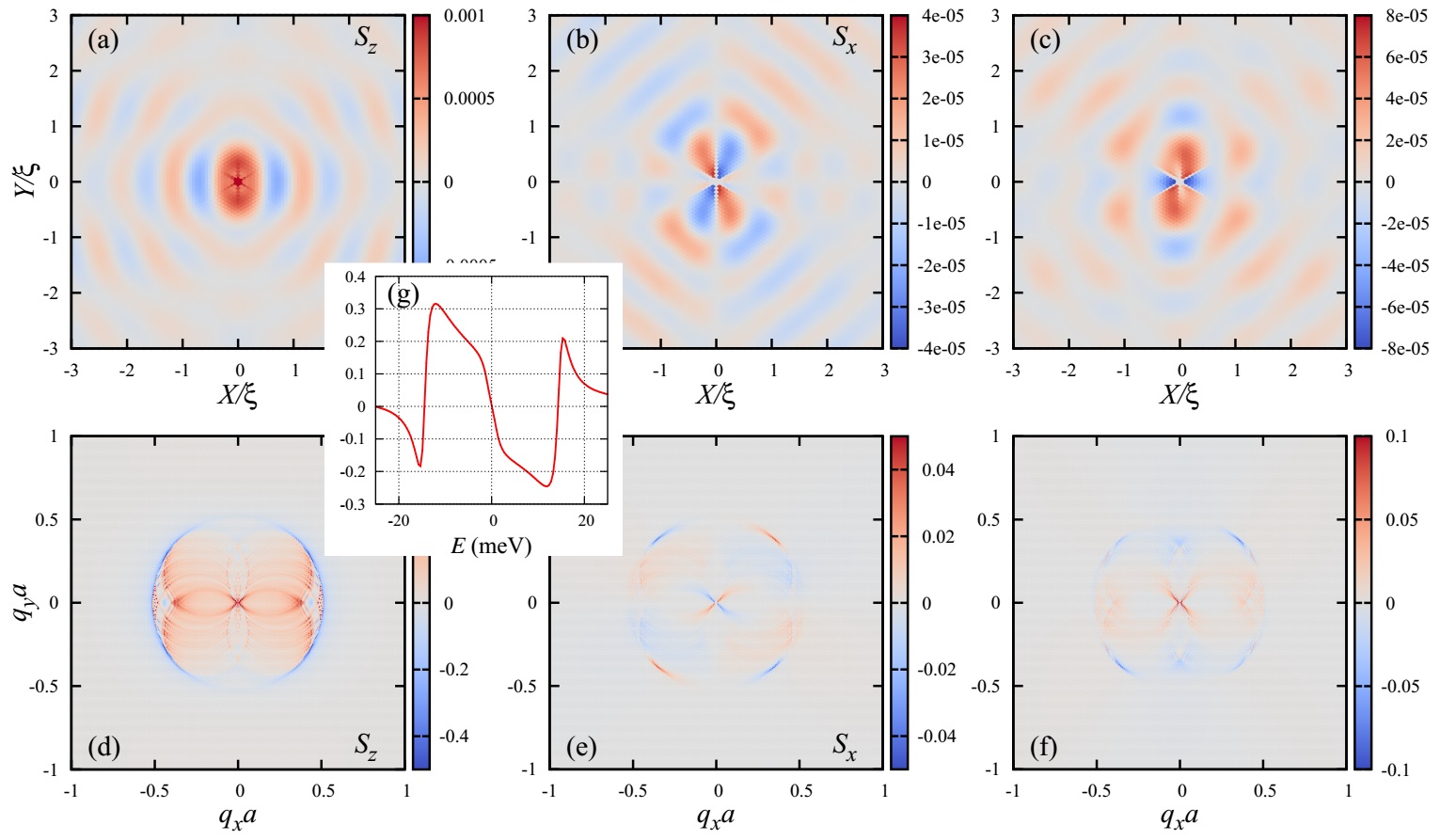}
	\caption{SP LDOSs of superconducting Bi$_2$Te$_3$ grown on FeTe$_{0.55}$Se$_{0.45}$ substrate with pairing potential $\Delta_{4a}$ and $z$-impurity, $\bm{J}=(0,0,J)$. The two in-gap resonance peaks are located at $E_{R}=\pm12$ meV. Parameters for numerical calculation are chosen as the same as Fig. \ref{fig2}. 
	}
	\label{fig3}
\end{figure*}

\begin{figure*}[tb]
	\centering
	\includegraphics[width=0.91\textwidth]{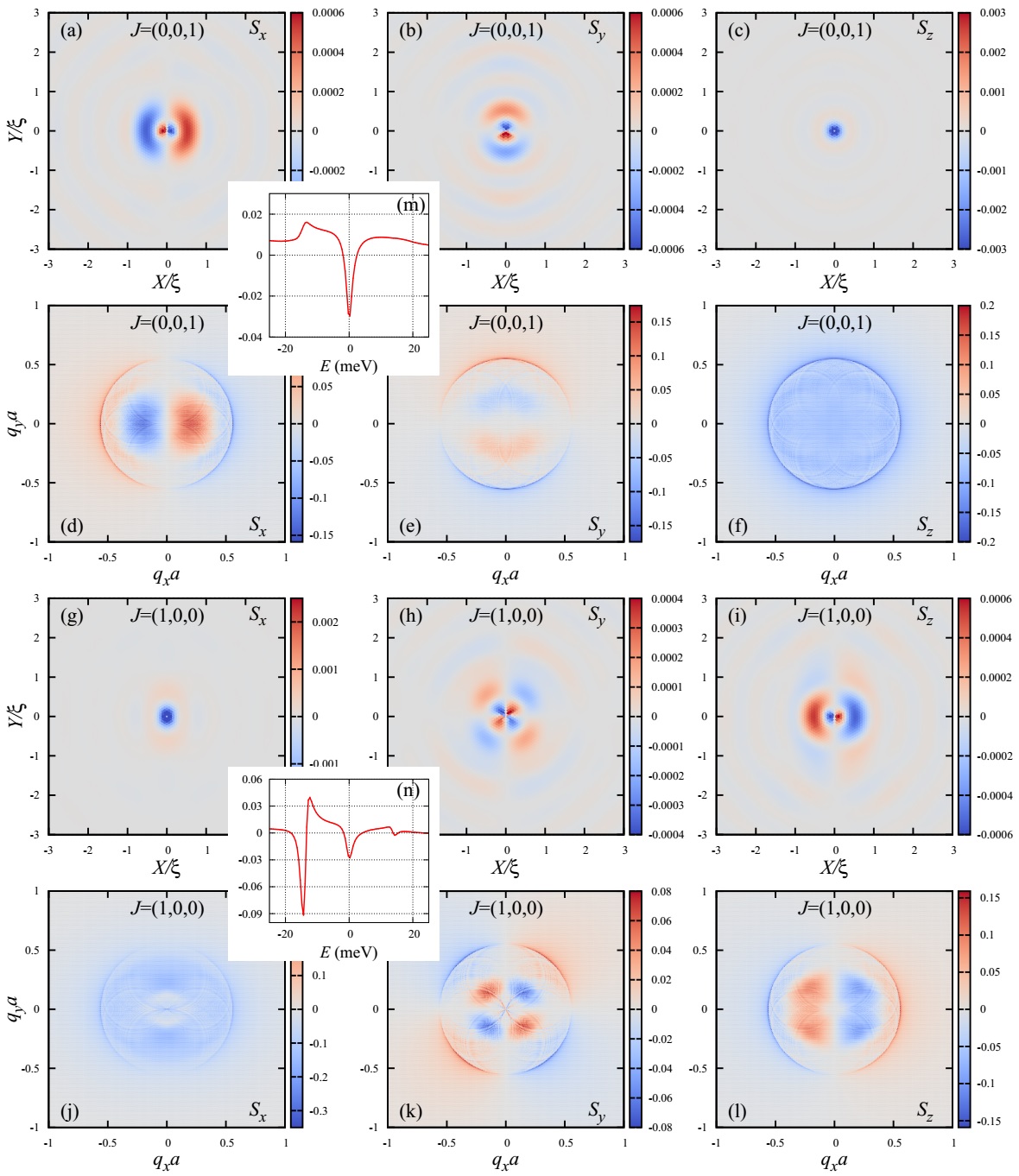}
	\caption{SP LDOSs for magnetic impurity located on the surface of Cu$_x$Bi$_2$Se$_3$ with pairing potential $\Delta_{4a}$. The upper and lower two rows show the results for magnetic $z$-impurity ($\bm{J}=(0,0,J)$) and $x$-impurity ($\bm{J}=(J,0,0)$), respectively. The first and third rows present SP LDOSs in real space. The second and forth rows present SP LDOSs after the Fourier transformation. The inserts (m) and (n) show the integration of $S_{z}(E,\bm{R})$ and $S_{x}(E,\bm{R})$ over $\bm{R}$, respectively. AFM resonance peaks can be found at $E=0$ for both these two cases. 
	}
	\label{fig4}
\end{figure*}

The eigenvalue of twofold rotation ($\mathscr{C}_{y}^{\prime}$) and inversion ($\mathscr{I}$) operations are both $\pm1$. For a given pairing potential which is parity-odd under these operations and the eigenvalue is $-1$, i.e., $\Delta{\xLongrightarrow{\mathscr{C}_{y}^{\prime} {\rm ~or~} \mathscr{I}}}-\Delta$, we find that, the results (\ref{eq23})-(\ref{eq25}) and (\ref{eq27})-(\ref{eq29}) are still correct. For the two species of pairing potentials belong to the $E_u$ representation of $D_3$ point group, we find the following relationships, $\Delta_{4a}\xLongrightarrow{\mathscr{C}_{y}^{\prime}}-\Delta_{4a}$, $\Delta_{4b}\xLongrightarrow{\mathscr{C}_{y}^{\prime}}+\Delta_{4b}$, $\Delta_{4a}\xLongrightarrow{\mathscr{I}}-\Delta_{4a}$, $\Delta_{4b}\xLongrightarrow{\mathscr{I}}-\Delta_{4b}$. Therefor, the results (\ref{eq23})-(\ref{eq25}) and (\ref{eq27})-(\ref{eq29}) will be manifested if the magnetic impurity is properly oriented and the pairing potential is nematic. However, if the pairing potential is a finite combination of $\Delta_{4a}$ and $\Delta_{4b}$, e.g., the chiral pairing proposed in \cite{WuF2017PRB}, the twofold rotation symmetry and the inversion symmetry are generally broken and the pattern preserves Eqs. (\ref{eq23})-(\ref{eq25}) and (\ref{eq27})-(\ref{eq29}) will not be observed. On the other hand, the nematic pairing $\Delta_{4a}$ or $\Delta_{4b}$ breaks the threefold rotation symmetry straightforwardly, so the symmetry of SP LDOS given in Eqs. (\ref{eq31})-(\ref{eq33}) will be violated for $\Delta_{4a}$ and $\Delta_{4b}$.

\section{Numerical results}\label{sec3}
It is difficult to get the analytic expressions of the SP LDOSs for the Hamiltonians given in Eqs. (\ref{eq1})-(\ref{eq3}). In order to get a more intuitive understanding and show the quantitative results of the SP LDOS influenced by the pairing symmetry, in this section, we present the numerical results of SP LDOS for the parameters given in Table \ref{tab1}. Before immersing into the detailed discussion of the numerical results, here we give some more details of the numerical calculations. The imaginary part of retarded Green's function has been chosen to be $0.001$eV. A mesh size with $1024\times1024$ points has been chosen for the SP LDOS and QPI calculations. In the evaluation of SP LDOS versus energy shown in Figs. \ref{fig2}(g) and \ref{fig3}(g), a mesh size up to  $4096\times4096$ points has been chosen to convergence the results. The fast Fourier transformation is applied to speed up the numerical calculation.

Firstly, we consider the 2-dimensional situation, few layer Bi$_2$Te$_3$ grown on FeTe$_{0.55}$Se$_{0.45}$ substrate. The four (twofold degenerated) eigenenergies of the BdG Hamiltonian, Eq. (\ref{eq4}), are,
\begin{eqnarray}
	& &E(\bm{k})=\pm\sqrt{\alpha\pm{2\sqrt{\beta}}}, \\
	\textrm{where~~~}\alpha&=&\gamma+\Delta_{\#}^2-(\varepsilon_{2d}-E_F)^2, \\
	\beta&=&\gamma(\varepsilon_{2d}-E_F)^2+\Delta_{\#}^2(M_{2d}^2+A_0^2c_{\#}^2), \\
	\gamma&=&M_{2d}^2+A_0^2(c_x^2+c_y^2)+(d_1^2R_1^2+d_2^2R_2^2),~~~~
\end{eqnarray}
$E_F$ is the Fermi energy, $\Delta_{\#}$ ($c_{\#}$) equals to $\Delta_{a}$ ($c_y(\bm{k})$) and $\Delta_{b}$ ($c_x(\bm{k})$) for $\Delta_{4a}$ and $\Delta_{4b}$ pairings, respectively. For each pairing potential ($\Delta_{4a}$ and $\Delta_{4b}$), the eigenenergies close to the Fermi surface have two nodes. Numerical calculation finds that these nodes are located at $\bm{k}a\approx(\pm0.26,0)$ and $(0,\pm0.26)$ for $\Delta_{4a}$ and $\Delta_{4b}$ pairings, respectively. These nodes make the density of states (DOS) to be V-shape like (see Fig. \ref{fig1}(a) for details), which is consistent with the experimental results shown in Ref. \cite{ChenM2018SciAdv} for few quintuple layers of Bi$_2$Te$_3$ grown on FeTe$_{0.55}$Se$_{0.45}$ substrate.

Now we analysis the QPI shown in Fig. \ref{fig1}. In order to get a clear representation, the QPI is plotted in a small region near the $\Gamma$ point. When the energy is located on the Fermi surface, i.e. $E=0$ shown in Fig. \ref{fig1}(d), the QPI manifests strong anisotropy along the $x$-direction, which indicates that the pairing potential $\Delta_{4a}$ has two nodes on the $k_x$-axis. Typical momentum transfer $\bm{q}a\approx(\pm0.52,0)$ shown in Fig. \ref{fig1}(d) is consistent with the position of nodes $\bm{k}a\approx(\pm0.26,0)$. When the energy is tuned away from the superconducting gap, i.e., $E=-0.03$eV and $0.03$eV shown in Figs. \ref{fig1}(c) and (f), QPI reveals sixfold rotation symmetry, which reflects the fact that the lattice structure is threefold rotation symmetric along the $z$-axis.
%Our result is a bit different from the previous study based on another tight-binding model {\color{blue}[Phys. Rev. B 98, 054502]}. The parameters used in this work is relevant to the practical material with small Fermi surface. We find that the two energy nodes are close to the $\Gamma$-point in the Brillouin zone, large momentum transfers shown in {\color{blue}[Phys. Rev. B 98, 054502]} are not allowed, which gives a clean representation of the threefold rotation symmetry broken.

%We have also calculated QPI for another pairing potential belongs to the $E_u$ representation of the $D_3$ point group, $\Delta_{4b}$, we find similar results (not presented here).

Next we study the SP LDOS for a magnetic impurity with spin oriented at $x$-direction, $\bm{J}=(J,0,0)$. Here, the nematic pairing symmetry $\Delta_{4a}$ are chosen for study. Fig. \ref{fig2}(g) shows the spatially averaged SP LDOS which is defined as $S_{x}(E)=S_{x}(E,\bm{q}=0)$. From this plot, one can find two in-gap Yu-Shiba-Rusinov (YSR) resonance states at $E\approx\pm{1.05}$meV. For the negative (positive) resonance energy $E=-1.05$meV ($E=1.05$meV), the resonance state and the magnetic impurity is ferromagnetic (antiferromagnetic) coupled. Figs. \ref{fig2}(a)-(c) show the spatial distributions of the YSR resonance states for different spin orientations respectively. From these plots, one can find that the SP LDOS exist Friedel oscillations with oscillation period on the scale of coherence length. The constraints from twofold rotation symmetry and inversion symmetry are manifested in these plots: (1) $S_x$ is invariant under the twofold rotation operation $\mathscr{C}_{y}^{\prime}$,  $(x,y)\rightarrow(-x,y)$, and the inversion operation $\mathscr{I}$, $(x,y)\rightarrow(-x,-y)$; (2) both $S_{y}$ and $S_{z}$ change signs after the twofold rotation operation $\mathscr{C}_{y}^{\prime}$; and (3) $S_{y}$ and $S_{z}$ are invariant under the inversion operation $\mathscr{I}$. The combination of twofold rotation and inversion operations give another strong constraint that $S_y$ and $S_z$ should be vanishing on the $x$- and $y$-axes. These feathers are more significant after the Fourier transformation, the numerical results are presented in Figs. \ref{fig2}(d)-(f), respectively.
%Our results are different from the previous studies on traditional $s$-wave and $p$-wave pairing.

Next we study the SP LDOS for another orientation of the magnetic impurity, $\bm{J}=(0,0,J)$. The spatially averaged SP LDOS $S_z(E)=S_z(E,\bm{q}=0)$ as a function of energy shown in Fig. \ref{fig3}(g) is qualitatively consistent with that given in Fig. \ref{fig2}(g): there are two in-gap YSR resonance peaks at $E\approx\pm{12}$meV respectively; for the negative (positive) resonance energy, the YSR state is ferromagnetic (antiferromagnetic) coupled to the impurity. Now we study the spacial distribution of the YSR resonance state with negative resonance energy, $E\approx-12$meV. Figs. \ref{fig3}(a)-(c) show the contour plots of SP LDOS of $S_z(E,\bm{R})$, $S_x(E,\bm{R})$ and $\left(-\frac{1}{2}S_x+\frac{\sqrt{3}}{2}S_y\right)(E,\bm{R})$, respectively. Generally, as discussed in the previous section, if the pairing potential preserves the threefold rotation symmetry, $S_z$ should be invariant under an 120$\degree$ rotation operation. However, both the real space SP LDOS \ref{fig3}(a) and its Fourier transformation \ref{fig3}(d) show a clear 180$\degree$ rotation symmetry. The threefold rotation symmetry broken is evident. Furthermore, as shown in Eq. (\ref{eq32}), if the pairing potential is invariant under the threefold rotation operation, plot \ref{fig3}(b) will coincide to \ref{fig3}(c) after an 120$\degree$ rotation operation. It is significant that these two plots can not coincide to each other. These observations demonstrate that SP LDOS can be applied to identify the threefold rotation symmetry broken if the pairing potential is nematic.

Finally, we study the SP LDOS for a magnetic impurity located on the surface of Cu$_x$Bi$_2$Se$_3$. Pairing potential has been chosen to be $\Delta_{4a}$ with pairing strength $\Delta_a=0.02$eV. The numerical results are presented in Fig. \ref{fig4}. As investigated in previous studies, due to the nontrivial topology of the pairing potential, there exists robust surface Andreev boundary states (SABSs) in the superconducting gap. Figs. \ref{fig4}(m) and (n) show the spacial averaged SP LDOS, $S_z(E)$ and $S_x(E)$ versus energy, for magnetic $z$-impurity and $x$-impurity respectively. In these plots, we find that the zero-energy SABSs strongly coupled to the magnetic impurity antiferromagnetically. As shown in Fig. \ref{fig4}(n), for the $x$-impurity, the spacial averaged SP LDOS is complicated. In addition to the antiferromagnetic (AFM) YSR resonance peak at $E=0$, there are two other resonance peaks near the band edge $E=15$meV. Here we focus on the zero-energy YSR resonance states. The first and third rows show the spacial distributions of the zero-energy YSR resonance states induced by $z$-impurity and $x$-impurity, respectively. The second and forth rows present the Fourier transformation of these SP LDOSs correspondingly. Inversion symmetry, $(x,y,z)\rightarrow(-x,-y,-z)$, is spontaneously broken for the surface impurity problem. The spatial distributions of the resonance states are dramatically different from the 2-dimensional case. One can find that Figs. \ref{fig4}(a), (b) and (i) are almost antisymmetric under the surface inversion operation $(x,y)\rightarrow(-x,-y)$. Numerical investigations show that the Fourier transformation of these SP LDOSs have both real and imaginary parts; the real parts are about two magnitudes smaller than the imaginary parts, respectively for each sub-figure. Here we give only the imaginary parts of SP LDOSs, which are plotted in Figs. \ref{fig4}(d), (e) and (l). Due to the intrinsic inversion symmetry broken of the surface states, it is difficult to identify the inversion symmetry of the bulk pairing potentials. Here we focus on the threefold symmetry broken of pairing potential. If the pairing potential and the impurity orientation is threefold rotation symmetric along the $z$-axis, i.e., $\bm{J}=(0,0,J)$, $S_z(E,\bm{R})$ and its Fourier transformation should be threefold rotation symmetric. One can find from the Fourier transformation Fig. \ref{fig4}(f) that $S_z(E,\bm{q})$ reveals twofold rotation symmetry. This symmetry broken is not as obvious as the 2-dimensional case. This may be induced by the topological surface states (TSSs). For the 3-dimensional material with a small Fermi surface, the TSSs of the parent material Bi$_2$Se$_3$ coexist with the SABSs \cite{WrayLA2010NatPhys,HaoL2011PRB,HaoL2015JPCM}. Generally, there are three kinds of scatterings: the scattering between TSSs, the scattering between TSSs and SABSs, and the scattering between SABSs. The first scattering preserves the threefold rotation symmetry. Only the scattering between SABSs can significantly reveal the symmetry broken if the pairing potential is nematic. This signature may be weakened by the inconsequential scatterings.

\section{Conclusion}\label{sec4}

We have analyzed the SPS of magnetic impurity resonance states for 2- and 3-dimensional nematic superconductors, in particular, few layer Bi$_2$Se$_3$ grown on FeTe$_{0.55}$Se$_{0.45}$ (2D) and Cu$_x$Bi$_2$Se$_3$ (3D) respectively. We focus on the relationship between nematic pairing symmetry and the topograph of SP LDOS. Different constraints from different symmetries are investigated, i.e., the SP LDOSs and their Fourier transformations should satisfy some constraints if the pairing potential preserves inversion symmetry, twofold rotation symmetry along the $y$-axis, or threefold rotation symmetry along the $z$-axis. The spacial distributions and the Fourier transformations of the SP LDOSs of YSR resonance states are investigated. For 2-dimensional case (few layer Bi$_2$Se$_3$ grown on FeTe$_{0.55}$Se$_{0.45}$), the SP LDOSs and their Fourier transformations reveal significant twofold rotation symmetry, threefold rotation symmetry is evident broken. For 3-dimensional case (Cu$_x$Bi$_2$Se$_3$), threefold rotation symmetry breaking can be identified by studying the Fourier transformation of the SP LDOS $S_z(E,\bm{R})$ for $z$-impurity orientation. In our studies, we use the tight-binding model with realistic parameters fitting the first principle calculations, our results can be tested experimentally.

In this work, we present only the numerical results for one species of the $E_u$ representation of the nematic pairing potential, $\Delta=\Delta_{4a}$. For the 2-dimensional case, spatial distribution and corresponding Fourier transformation of resonance states with only negative resonance energy are presented. The numerical calculations for other situations have been accomplished, i.e., pairing potential is chosen to be $\Delta_{4b}$ and the SP LDOS of resonance states with positive resonance energy. We did not find qualitative difference, so the numerical results are not presented here.

\section*{Acknowledgment}\label{secAck}
We appreciate support from the NSFC under Grants No. 11504106 and No. 11447167 and the Fundamental Research Funds for the Central Universities under Grant No. 2018MS049.

% Specify following sections are appendices. Use \appendix* if there only one appendix.

%\appendix
%

%-----------------------------------------------------------------
% Sec**: References
%-----------------------------------------------------------------
%\nocite{*}
%\bibliographystyle{apsrev4-1}
%\bibliography{magImp}

%

\end{document}